\documentclass[iop]{emulateapj}
\usepackage{graphicx,aas_macros,url}
\shorttitle{CXOCY J220132.8-320144}
\shortauthors{Chen et al.}

\begin{document}

\title{Missing Lensed Images and the Galaxy Disk Mass in CXOCY J220132.8-320144}
\author{Jacqueline Chen\altaffilmark{1}, Samuel K. Lee\altaffilmark{1,2}, Francisco-Javier Castander\altaffilmark{3}, Jos\'{e} Maza\altaffilmark{4}, and Paul L. Schechter\altaffilmark{1}}
\altaffiltext{1}{Kavli Institute for Astrophysics and Space Research, Department of Physics, Massachusetts Institute of Technology, \\
77 Massachusetts Avenue, Cambridge, MA 02139}
\altaffiltext{2}{Princeton Center for Theoretical Science, Princeton University, Princeton, NJ 08544}
\altaffiltext{3}{Institut de Ci\`{e}ncies de lÕEspai (IEEC-CSIC), E-08193 Bellaterra (Barcelona), Spain}
\altaffiltext{4}{Departamento de Astronomia, Universidad de Chile, Casilla 36-D Santiago, Chile}

\begin{abstract}
The CXOCY J220132.8-320144 system consists of an edge-on spiral galaxy lensing a background quasar into two bright images.  Previous efforts to constrain the mass distribution in the galaxy have suggested that at least one additional image must be present \citep{castander_etal06}.  These extra images may be hidden behind the disk which features a prominent dust lane.  We present and analyze Hubble Space Telescope (HST) observations of the system.  We do not detect any extra images, but the observations further narrow the observable parameters of the lens system.  We explore a range of models to describe the mass distribution in the system and find that a variety of acceptable model fits exist.  All plausible models require 2 magnitudes of dust extinction in order to obscure extra images from detection, and some models may require an offset between the center of the galaxy and the center of the dark matter halo of 1 kiloparsec.  Currently unobserved images will be detectable by future James Webb Space Telescope (JWST) observations and will provide strict constraints on the fraction of mass in the disk.
\end{abstract}


\section{Introduction}
\label{sec:intro}

Gravitational lensing is a potent probe of the matter distribution in the central regions of lensing galaxies.  A particularly promising application of gravitational lensing is to the study of the mass distribution in edge-on disk galaxies.  As the geometry of the luminous disk galaxy differs significantly from the dark matter halo in which the galaxy is embedded, disentangling the relative contributions to the total mass of the luminous and the dark components is, at least in principle, feasible. 

The decomposition of the disk galaxy from the dark matter halo has customarily employed a maximum disk model \citep{vanalbada_etal85}.  In systems where the mass-to-light ratio of the disk is unknown, the contribution of the disk mass to the rotation curve is taken to be as large as permitted by the observed rotation curve, and the galaxy dominates the mass near the center of the galaxy.  The accuracy of the maximum disk model is unsettled, although \citet{debattista_sellwood98} suggest that disk galaxies with bars are likely to be maximal.  The characteristic radius at which one typically characterizes the fraction of matter in disk galaxy is 2.2 disk scale lengths ($2.2R_d$), as this is the radius at which the circular speed peaks for a thin exponential disk.  In what follows, we parameterize the size of the galaxy disk by the fraction of mass in the disk galaxy inside a sphere of radius $2.2R_d$, referred to hereafter as $f_{disk}(2.2R_d)$.

Progress on this front has until now been impeded by the small number of edge-on spiral galaxy lenses.  Four lens systems with quasar sources are known:  Q2237+0305 \citep{huchra_etal85}, B1600+434 \citep{jackson_etal95,jaunsen_hjorth97}, PMN J2004-1349 \citep{winn_etal03}, and CXOCY J220132.8-320144 \citep{castander_etal06}.  In addition the Sloan WFC Edge-on Late-type Lens Survey \citep[SWELLS,][]{treu_etal11} has assembled 20 galaxy-galaxy lenses.  

As the lens galaxy in Q2237+0305 has a low redshift, $z_{\rm lens} = 0.04$, the lensing is mostly sensitive to the bulge mass  \citep{vandenven_etal10}.  Still, combining with stellar kinematic information, \citet{trott_webster02} and \citet{trott_etal10} find that the lens galaxy has a submaximal disk -- i.e., the disk {\it does not} dominate the mass in the inner regions of the galaxy.  \citet{koopmans_etal98} and \citet{maller_etal00} put constraints on a lower-limit for the halo axis ratio in B1600+434, finding that is greater than $q \sim0.5$, and \citet{maller_etal00} rules out a maximal disk.  For PMN J2004-1349, only the bulge-to-disk mass is constrained \citep{winn_etal03}.

The SWELLS survey team has analyzed one galaxy-galaxy lens, SDSS J2141-0001, combining gas and stellar kinematics data with the constraints from the lensed arc of the source galaxy \citep{dutton_etal11,barnabe_etal12}.  In one analysis, performed before stellar kinematic data became available, the disk of the lensing galaxy was found to be submaximal, and $f_{disk}(2.2R_d)$ is 0.45.  In the analysis including stellar kinematic data, the disk of the lensing galaxy is maximal and $f_{disk}(2.2R_d)$  is 0.72.

CXOCY J220132.8-320144 (hereafter CX 2201) was discovered as part of the Cal\'{a}n-Yale Deep
Extragalactic Research (CYDER) survey of {\it Chandra} fields \citep{castander_etal03,treister_etal05}.  Ground-based
optical imaging and spectroscopy were carried out, confirming the lensed
nature of the system and measuring $z = 0.323$ and $z = 3.903$ for the
galaxy and the quasar respectively \citep{castander_etal06}.  At first examination, CX 2201 may not appear to be a prime candidate for studying edge-on spiral galaxy lenses, since it has only two observed images of the background quasar.  It displays some intriguing features, however, that make it worthy of inquiry.  The lensing galaxy has no visible bulge, further eliminating one possible impediment to a clean estimate of the dark matter to disk mass fraction.  In addition, both lensed images lie close to the plane of the disk and have similar fluxes.  If the galaxy disk were maximal, the matter distribution should be highly elongated and it might be expected that  CX 2201 would be a four-image system.  Previous efforts to constrain the mass distribution in the galaxy have suggested that at least one additional image must be present \citep{castander_etal06}.  These extra images may be hidden behind the disk which features a prominent dust lane.  

In this paper, we present Hubble Space Telescope (HST) observations of the system.  These were carried out at longer wavelengths than the original detection observations in an effort to observe extra images that may have been extincted by dust in the disk at shorter wavelengths.  In addition, we improve upon the astrometry and photometry of the observed images and use the improved constraints to narrow a range of lensing models to describe the mass distribution in the system.

\section{HST Observations}
\label{sec:obs}

\begin{figure}
\includegraphics[width=3.3in]{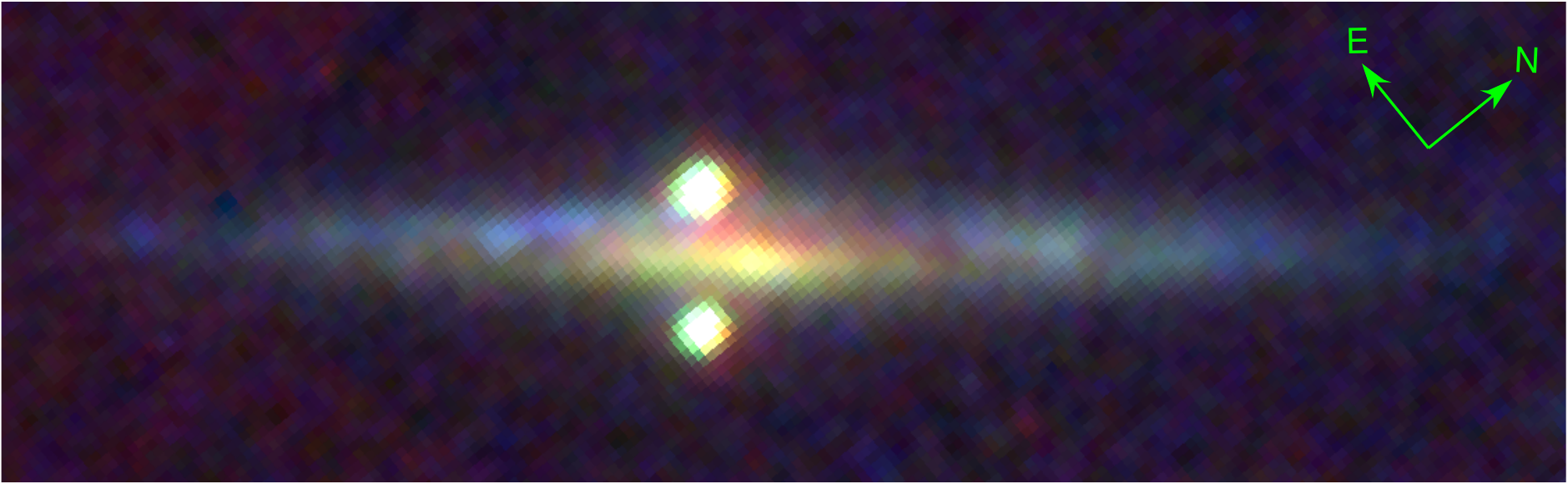} \caption{Combined 160W, 475W, and 814W HST image of CX 2201.  Image A is located above the disk of the galaxy and image B is located below.  They are separated by
 $\sim$$0\farcs 8$.  The size of the displayed image area is $9\arcsec \times 2\farcs7$.  
\label{fig:Fig1}}
\end{figure}

Observations using the HST
Advanced Camera for Surveys (ACS) and the Near-Infrared Camera and
Multi-Object Spectrometer (NICMOS) were carried out on 2006 May 11 and
2006 April 13, respectively (GO:10518). ACS observations were taken in the F814W and
F475W filters; NICMOS observations were taken in the F160W filter.

ACS imaging consisted of 4 dithered exposures taken in ACCUM mode in
each filter; total exposure times were 2228 seconds and 2252 seconds
in the F814W and F475W filters, respectively.  The exposures were
reduced and calibrated using the CALACS calibration pipeline
(including the PyDrizzle algorithm).  NICMOS imaging consisted of 4 dithered exposures taken in MULTIACCUM mode;
total exposure time was 2688 seconds.  The exposures were reduced and
calibrated using the CALNICA/CALNICB calibration pipeline.  Quadrant bias
was corrected using the \textit{pedsub} routine in the STSDAS IRAF
package.  

In each image, the two quasar images straddle the galaxy, with the line
connecting the two images intersecting the galaxy disk at a point offset from
the galaxy center.  There appears to be no pronounced bulge component in
the galaxy.  The disk is highly-inclined, but has a non-zero projected
axis ratio and is thus not perfectly edge-on. A dust lane is visible in
the ACS images.  Figure \ref{fig:Fig1} shows a combined 160W, 475W, and 814W HST image of CX 2201.

The images were fit using the GALFIT program \citep{peng_etal02} to extract positions and
fluxes of the images and the galaxy and other parameters necessary to
completely characterize the galaxy. The galaxy was modeled as an
exponential disk, while the quasar images were modeled as Gaussians (quasi-delta functions).  The
positions, fluxes, and scale radii of the quasar images and the galaxy
were allowed to vary, as were the galaxy position angle and axis ratio.
The combined model was convolved with the point spread functions (PSFs) of
each instrument, which were calculated using the {\sc tinytim} program.  In
constructing the PSFs, a frequency power-law of $-0.5$ was assumed for the
source quasar spectrum.

\begin{table}
\begin{center}
\caption{CX 2201 Lens Model Constraints} \label{table:Tab1}
\begin{tabular}{cccc}
    \hline
    \hline
Parameter            &      Image A      &      Image B      &  Galaxy \\
    \hline
x position (as)$^{a}$      &      -0.225       &     -0.210       &  0.000 \\
y position (as)      &       0.353        &      -0.468       &  0.000 \\
position error (mas) &       0.44        &       0.45        &  3.57  \\
relative flux$^{b}$
                     &     1.00          &     0.72          &  \\
axis ratio           &            &            &   0.12  \\
disk scale length, $R_d$ (as)$^{c}$    &            &            &   1.60  \\
    \hline
\end{tabular}
\end{center}
$^{a}${Image positions are calculated in the rotated frame where the
galaxy plane is horizontally aligned and both quasar images are to the left of the
galaxy center.}\\
$^{b}${Relative flux in the F160W filter.}   \\
$^{c}${$R_d$ increases with
decreasing filter wavelength in the three images, as expected for a spiral
galaxy.  $R_d$ is $1\farcs01$, $1\farcs57$, and $2\farcs37$ in the F160W, F814W, and F475
filters, respectively.} 
\end{table}

The dust lane is quite prominent in the ACS images, and is visible even in the
NICMOS image.  Despite the difficulty of fitting with the dust lane obscuring much of the
galaxy center, the positions of the quasar images and the galaxy are
consistent in the different filters.  The results of the best fit are
shown in Table~\ref{table:Tab1}, where the error-weighted average values are given and positions are rotated such that the
galaxy plane is horizontally aligned.  The position angle of the galaxy East of North is $140.5\degr$, $139.3\degr$, and $140.1\degr$ in the F160W, F814W, and F475
filters, respectively.  In every filter observed, the quasar images do not straddle the center of the galaxy;  instead, they are offset from the center of the galaxy and lie left of the
galaxy center in the rotated frame.  This offset is a key factor in the mass models used to describe the lens system and is discussed further in Section \ref{sec:mod}.  We measure an offset along the disk plane from the galaxy center to the line connecting the two images
of $\sim$$0\farcs2$, which is smaller than the value of $\sim$$0\farcs3$ previously measured
by \citet{castander_etal06} using ground-based data.

Although GALFIT gives the same galaxy center position in all the
filters, the fit is quite uncertain.  Indeed, measuring the center
through other means (such as using elliptical galaxy isophotes or masking
out the dust lane in the fit) gives offsets which vary from $\sim$$0\farcs15$
to $\sim$$0\farcs5$.  In addition, the disk scale length, $R_d$, varied over the
different filters, increasing with decreasing wavelength.  This is
consistent with a spiral galaxy with redder old stars in the core and
bluer star-forming regions in the outer arms.  However, this variation
introduces additional uncertainty in defining $R_d$ for models and is discussed further in later sections.  

\begin{table}
\begin{center}
\caption{CX 2201  Photometry} \label{table:Tab2}
\begin{tabular}{cccc}
    \hline
    \hline
                  &      F160W        &      F814W        &      F475W        \\
    \hline
Image A           & $21.67 \pm 0.01$  & $23.46 \pm 0.01$  & $25.96 \pm 0.02$  \\
Image B           & $22.02 \pm 0.02$  & $23.76 \pm 0.01$  & $26.19 \pm 0.02$  \\
Galaxy            & $19.65 \pm 0.01$  & $21.19 \pm 0.01$  & $22.61 \pm 0.01$  \\
Flux Ratio   &      1.39         &      1.32         &      1.23      \\
(A/B) & & & \\
    \hline
\end{tabular}
\end{center}
\end{table}

The results of the photometry for each filter are shown in
Table~\ref{table:Tab2}.  The flux ratio A/B is higher in the F160W filter than in the
F814W and F475W filters; that is, image A appears slightly redder than
image B.  In addition, \citet{castander_etal06} previously measured a flux ratio A/B of $\sim$$1.1$ in the Sloan $r$-band with the Magellan Clay telescope.  Since image A is closer to the disk than image B, this suggests
differential extinction by a galaxy dust profile which decreases with
distance from the plane of the galaxy.  However, other possible
explanations for the apparent reddening include microlensing and intrinsic
variability on timescales greater than the time delay between the images.  Without an explicit accounting for those effects, we cannot determine an unreddened flux ratio.  In the lens modeling, we use the F160W flux ratio which would be the least affected by reddening, and microlensing and intrinsic variability effects are indirectly accounted for by increasing the error in the flux ratio measurement.  Thus, reddening effects are unlikely to bias our modeling results.  

\section{Gravitational Lens Modeling}
\label{sec:mod}

We now fit several mass models to the observations using the observational constraints summarized in Table \ref{table:Tab1}.  The two observed quasar images provide 4
position constraints and 1 flux ratio constraint, a total of 5.  For models that have additional unobserved lensed images, the non-detection of such images gives us additional constraints.  

For every lens model considered, all have common inputs.  Flux errors are
artificially increased to $20\%$ to allow for microlensing or intrinsic
variability; likewise, position errors are increased to 5 milliarcseconds
(mas) to account for substructure effects.  For every model we test, 2 of the free parameters must be used to fit the position of the source, $x_{s}$ and $y_{s}$.  A Gaussian prior on the position of the center of the lensing galaxy is included with the standard deviation given by the measured error in Table \ref{table:Tab1}.

Sampling the lens model parameter space is performed using {\tt emcee}, the Python ensemble sampling toolkit for affine-invariant MCMC \citep{foremanmackey_etal12}, in order to sample the parameter space of the mass models.  

\subsection{Two-Image Lens Models}

Under the assumption that the two observed lensed images are the only images created by the lens, the small number of observable constraints can overconstrain only simple lens models with 4 or fewer free parameters.  We choose for a simple lens model a singular isothermal ellipsoid (SIE).  A SIE is a logical model to test, as many strong lens systems with early-type lensing galaxies have been successfully modeled by mass profiles with the same slope as the SIE \citep{koopmans_etal06}.  

\begin{table}
\begin{center}
\caption{Lens Models} \label{tab:models}
\begin{tabular}{lll}
\hline
\hline
Model & Parameter & Description \\
\hline
SIE & $b$ & halo Einstein radius\\
 & $q$ & halo axis ratio\\
 & $x_s$ & source $x$ position \\
  & $y_s$ & source $y$ position\\
\hline
NIE + Chameleon & $b_h$& halo Einstein radius \\
& $q_h$ & halo axis ratio \\
& $s_h$ & halo core radius \\
& $b_g$ & galaxy Einstein radius \\
 & $x_s$ & source $x$ position \\
  & $y_s$ & source $y$ position\\
\hline
\end{tabular}
\end{center}
\end{table}

A SIE has a dimensionless projected surface density
\begin{equation}
\kappa(\xi) = \frac{b}{2 \xi},
\end{equation}
where $\xi^2 = x^2 + y^2/q^2$ and where $q$ is the axis ratio of the mass distribution.  In the case where the halo is spherical and $q=1$, $b$ is the Einstein radius and related to the 1-d velocity dispersion, $\sigma$, by  $b =  4 \pi (\sigma/c)^2 D_{\rm ls}/D_{\rm os}$, where $D_{\rm ls}$ and $D_{\rm os}$ are the angular diameter distances from the lens to the source and from the observer to the source, respectively.  Solutions for the deflection and magnification from this model are given by \citep{kassiola_kovner93,kormann_etal94}.  The model and free parameters are summarized in Table \ref{tab:models}.

Models of the two-image system with a SIE have a best-fit $\chi^2 \sim 1000$.  A summary of the results of our modeling runs is presented in Table \ref{tab:results}.  The poor outcomes are due to the asymmetry of the observed images.  Both images are located to the left of the galaxy center (in the rotated frame).  In addition, they have similar fluxes.  The SIE model creates solutions where one image is located to the left of the galaxy center and the counterimage (on the opposite side of the disk) is found to the right of the galaxy center and with a much smaller relative flux.

A simple solution to this problem would be to shift the galaxy center to be collinear with the observed images.  Such a shift might be justified.  The lensing galaxy in CX 2201 could be analogous to NGC 4631, a nearby edge-on spiral galaxy.  NGC 4631 is lopsided, the disk of the galaxy extending further from one side of the galaxy center than from other other side.  In CX2201, the dust lane in the galaxy obscures the center of the galaxy.  It is possible that the true center of CX 2201 lies offset from the midpoint of the galaxy disk.  However, NGC 4631 is interacting with another edge-on spiral neighbor, NGC 4656 \citep{roberts68,weliachew69,weliachew_etal78}, and CX 2201 has no such interacting neighbor.  In addition, the amount of dust obscuration should differ by wavelength, such that we might expect to measure different offsets along the disk plane from the galaxy center to the line connecting the two images, and we do not.  Every single filter in our data of CX 2201 shows that the images are $\sim0.2"$ away from the galaxy center, and Magellan observations reported in \citet{castander_etal06} show the images as $\sim0.3"$ away from the galaxy center.  On the whole, it is unlikely that dust lane obscuration is sufficient to motivate a galaxy center located between the images.   Adopting such a model for the sake of argument, however, achieves good model fits, $\chi^2=1.5$.

\begin{table*}
\begin{center}
\caption{Summary of Lens Modeling Results}
\begin{tabular}{c|l|c|l}
\# of Model Images & Model& Best-Fit $\chi^2$ & Description \\
\hline
\hline
&&&\\ [-1.7ex]
2 & SIE  & 1790 & \parbox{3in}{Model puts one image on the other side of the halo center instead of having both images on one side.  }\\
&&&\\ [-1.7ex]
\hline
&&&\\ [-1.7ex]
 2 & \parbox{1.5in}{SIE -- galaxy center offset} & 1.5 & \parbox{3in}{Galaxy center offset is not supported by observations.}  \\
&&&\\ [-1.7ex]
\hline
\hline
&&&\\ [-1.7ex]
3 &  \parbox{1.5in}{NIE + Chameleon -- no dust extinction}& 29 &  \parbox{3in}{$\chi^2$ for observed images:  $\chi^2_{2\rm im} = 3$.  $\chi^2$ for extra image:  $\chi^2_{\rm extra} = 26$.  Model parameters are not well constrained.  Observations of the extra image and a tighter constraint on the disk scale length are required to distinguish mass models.  }\\
&&&\\ [-1.7ex]
\hline
&&&\\ [-1.7ex]
3 &  \parbox{1.5in}{NIE + Chameleon -- 1 mag dust extinction}& 12 &  \parbox{3in}{$\chi^2_{2\rm im} = 0.8$.  $\chi^2_{\rm extra} = 11$.  }\\
&&&\\ [-1.7ex]
\hline
&&&\\ [-1.7ex]
3 &  \parbox{1.5in}{NIE + Chameleon -- 2 mag dust extinction}& 0.5 &  \parbox{3in}{$\chi^2_{2\rm im} = 0.4$.  $\chi^2_{\rm extra} = 0$.  }\\
&&&\\ [-1.7ex]
\hline
\hline
&&&\\ [-1.7ex]
5 & \parbox{1.5in}{NIE + Chameleon -- 2 mag dust extinction} & 7 & \parbox{3in}{$\chi^2_{2\rm im} = 7$.  $\chi^2_{\rm extra} = 0$.  Unacceptable fits for observed image positions.} \\
&&&\\ [-1.7ex]
\hline
&&&\\ [-1.7ex]
5 & \parbox{1.5in}{NIE + Chameleon -- 2 mag dust extinction -- halo center offset } & 0.3 & \parbox{3in}{$\chi^2_{2\rm im} = 0.3$.  $\chi^2_{\rm extra} =0$.  } \\
&&&\\ [-1.7ex]
\hline
&&&\\ [-1.7ex]
5 & \parbox{1.5in}{NIE + Chameleon + $10^9 M_{\sun}$ substructure -- 2 mag dust extinction} & 0.5 & \parbox{3in}{Substructure mass is large; smaller masses do not improve fit over no substructure.} \\
&&&\\ [-1.7ex]
\hline
\hline
\end{tabular} 
\label{tab:results}
\end{center}
\end{table*}

\subsection{Models with 3 \& 5 Images}

A more realistic mass model would include both a component for the dark matter halo and a component to describe the disk galaxy.  We describe the dark matter by a non-singular isothermal ellipsoid (NIE), which is a SIE modified by the inclusion of a core radius, such that the dimensionless projected surface density is
\begin{equation}
\kappa(\xi) = \frac{b}{2 ~(s^2 + \xi^2)^{1/2}},
\end{equation}
where $s$ is the core radius.  The dark matter halo, then, has 3 free parameters associated with it:  $b_h$, $q_h$, and $s_h$.  A hard prior is included in the modeling such that $b_h>0$, $q_h = [0,1]$ and $s_h > 0$.

Simulations of dark matter halos find an inner mass profile of halos that is shallower than an isothermal profile \citep[e.g.,][]{navarro_etal97,moore_etal99a}.  On the other hand, in a halo with a galaxy in the center of it, lensing studies show that isothermal is a good fit for the total mass profile (dark matter + galaxy).  A NIE profile provides sufficient flexibility to capture the change in the density profile in the central regions and a range of possible dark matter profile behavior \citep[c.f.,][]{dutton_etal11}.  

The galaxy disk is described by an exponential profile.  We approximate the exponential profile by the difference of two NIEs, as described by \citet{maller_etal00}, where the core radii of the NIEs are given by $s_{g,1}=(1/3)R_d$ and $s_{g,2}=(7/3)R_d$.  This is referred to as a `Chameleon' profile.  Since the axis ratio and the scale length of the disk is measured observationally, only 1 free parameter is associated with the galaxy:  $b_g$.  A hard prior is included such that $b_g>0$.

With the addition of $x_s$ and $y_s$ to describe the source position, and our disk+halo model requires a total of 6 free parameters.  The model and free parameters are summarized in Table \ref{tab:models}.  With 6 free parameters, more than 2 images are needed to constrain the model.  It might be thought that every additional image should provide an additional 3 observable constraints.  This is incorrect for the constraints on non-detections of images.  There are no real constraints on the location of such images.  However, there are constraints on the brightness of extra images.  

\begin{figure}
\includegraphics[width=1.08in]{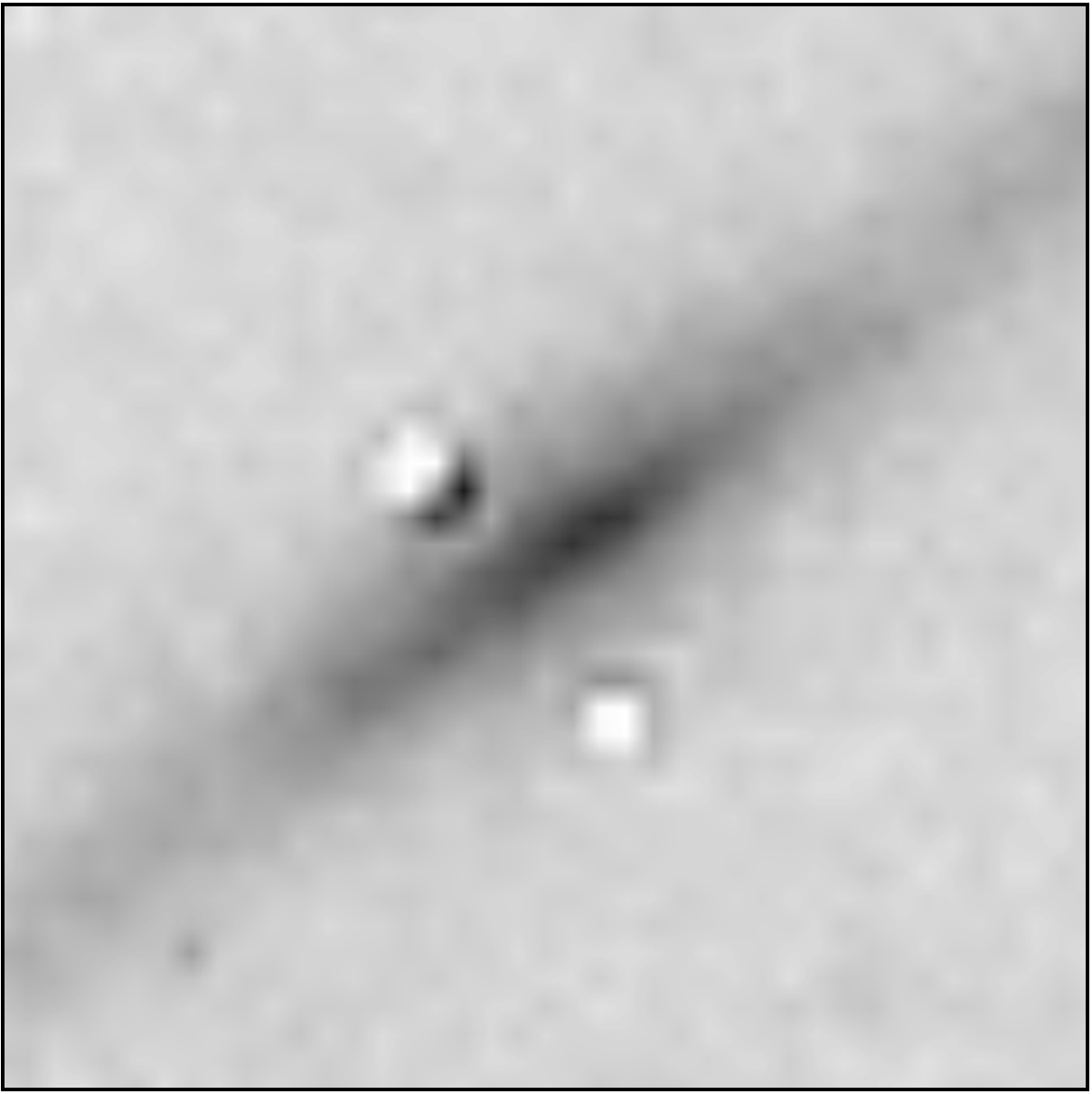} 
\includegraphics[width=1.08in]{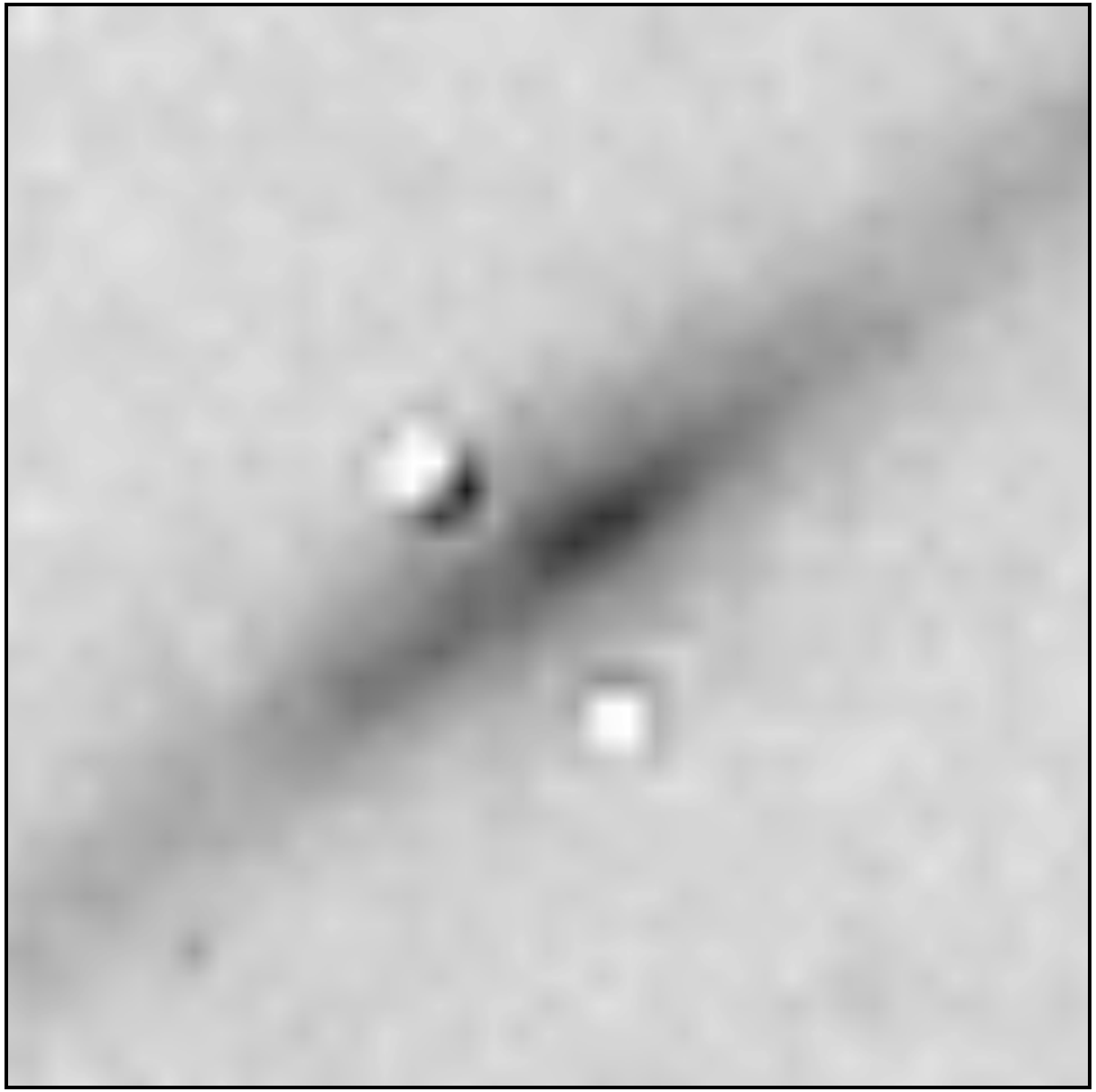} 
\includegraphics[width=1.08in]{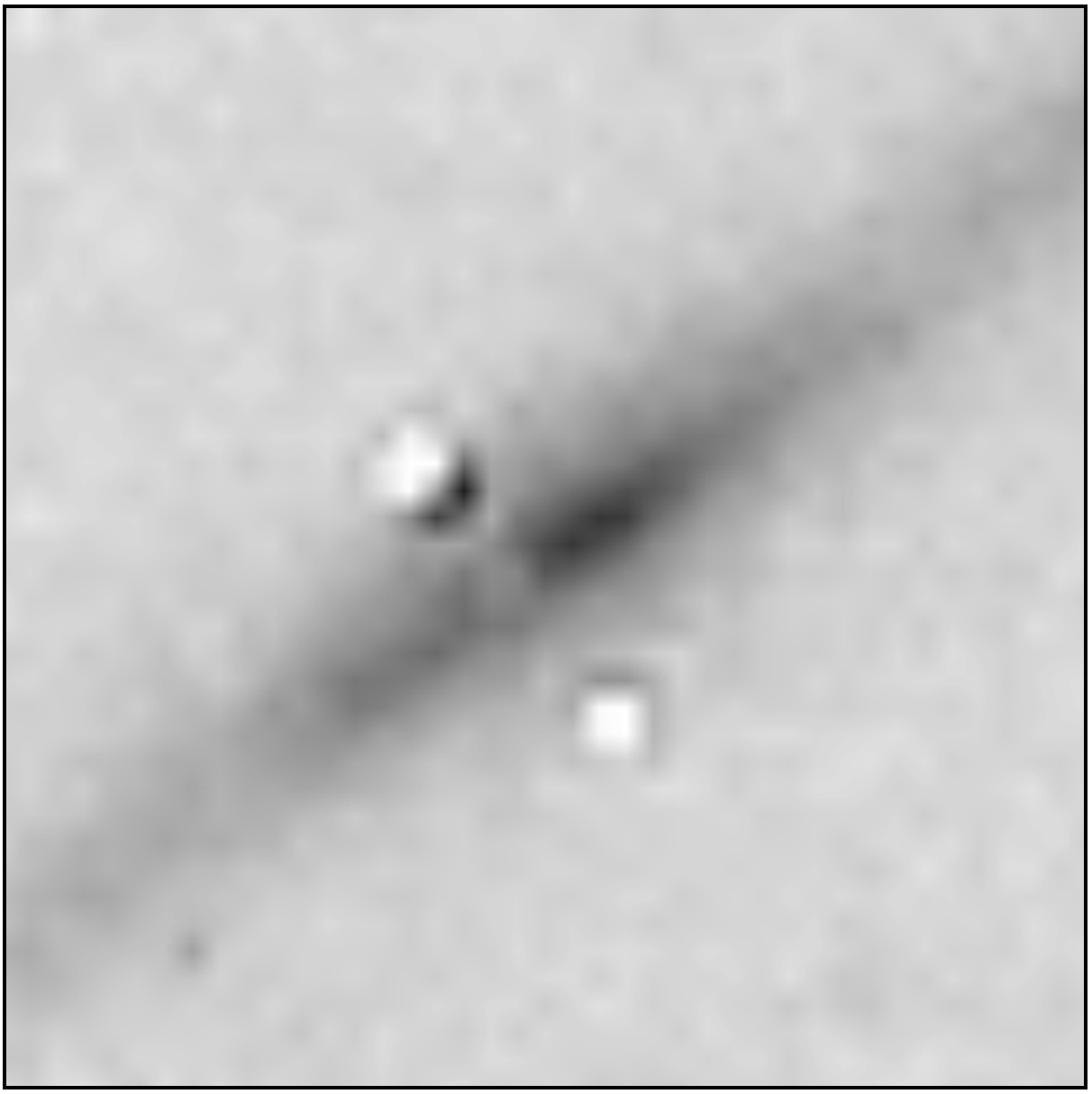} 
\includegraphics[width=1.08in]{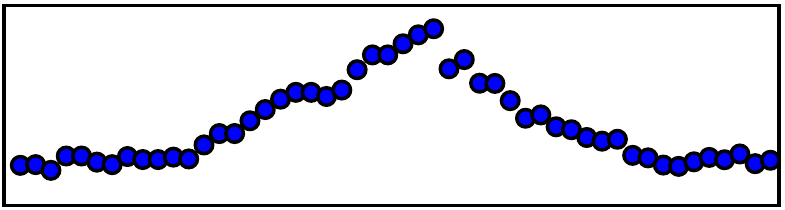} 
\includegraphics[width=1.08in]{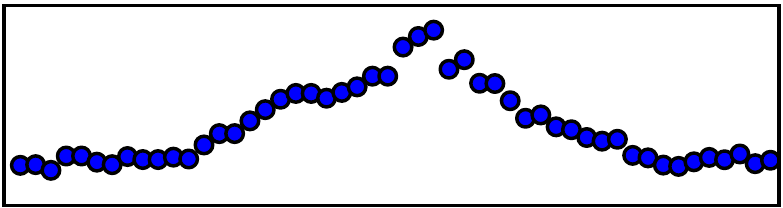} 
\includegraphics[width=1.08in]{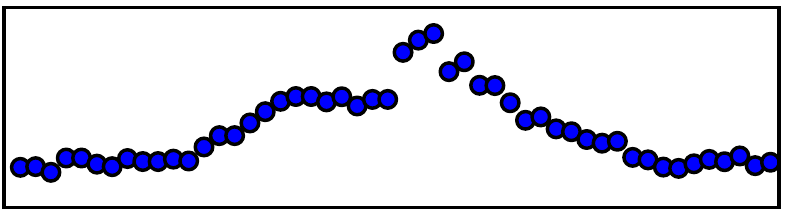} 
\caption{The top panels show the galaxy with images subtracted, while the bottom panels show the resulting brightness in a slice along the plane of the disk.  Left:  The two bright modeled images are subtracted from 160W observation.  Center:  An image modeled with flux $1/20$ the brightness of image A is subtracted from the left-hand image.  Right:  An image modeled with flux $1/10$th the brightness of image A is subtracted from the left-hand image.    
\label{fig:subtract}}
\end{figure}

In order to determine how bright an extra image needs to be to be detected, we model an extra image by smoothing the flux in the image over the measured point spread function (PSF) and locate it between the two observed images, where extra images are expected.  Model images with fluxes greater than $1/10$th the brightness of image A are clearly detectable in the observations.  Figure \ref{fig:subtract} shows that subtracting a model image of that brightness is observable as a "divot" in the galaxy disk.  We set the flux ratio of image A to an extra image of 10 (A/[extra image] $=10$) as the threshold of detectability and consider this a 2.5$\sigma$ detection.  Images of this brightness add an additional $\chi^2_{\rm extra} = 6.25$ to the $\chi^2$ of the model.  Images that are $N$ times brighter than the cutoff add $\chi^2_{\rm extra} = N \times 6.25$ to the model.

A model image could be significantly brighter than this cut-off and still be unobservable due to extinction by the dust lane in the galaxy.   By way of illustration, consider the dust extinction towards the center of the Milky Way.  We observe CX 2201 at a redshift of $z=0.323$ at $1.6\mu$m.  This corresponds to $J$ band ($1.2\mu$m) observations in the rest frame.  The dust extinction toward the center of the Milky Way is $A_V=31$ \citep{scoville_etal03} and $A_J=8$ \citep{rieke_lebofsky85}.  So significant amounts of dust extinction might be expected near the centers of edge-on spiral galaxies with dust lanes such as CX 2201.  As discussed in the following sections, all extra images lie in the plane of the disk.  For this reason, we test models which add a set amount of extinction to the modeled extra image or images.  In a model with 1 magnitude of extinction, extra images with a flux ratio of 4 are at the limit of detectability.  At 2 magnitudes of extinction, the limit is a flux ratio of 1.6, and at 3 magnitudes the limit is a flux ratio of 0.63.

\subsection{Models with One Extra Image}

Using our NIE + Chameleon model, two different lens system configurations could give rise to 3 image solutions:  a mass model without a naked cusp \citep[e.g.,][]{ibata_etal99,egami_etal00}  and a naked cusp mass model.  In a model without a naked cusp, any non-singular mass distribution should produce an odd number of lensed images.  Given this, observed lens system 2-image systems are truly 3 image systems, where the central image has been demagnified sufficiently to be unobservable.  The predominance of lens systems with even numbers of images has been used to limit the core radius in lensing systems, as small cores result in the demagnification of the central image \citep{keeton03b}.  In this configuration, the source lies between the inner diamond-shaped tangential caustic and the outer elliptical radial caustic, and the two outer images have different image parities.

Naked cusps occur in systems with highly elongated mass distributions, when the tangential caustic extends outside the radial caustic.  Generically, a source inside this extension produces three roughly collinear images \citep[e.g.,][]{keeton_kochanek98}.  It is commonly reported that the central image is of similar brightness to the outer images.  The only current detection of a naked cusp in a system with a galaxy lens is APM 08279+5255 \citep{lewis_etal02b}, where the central image is observed to be $0.2$ times the brightness of the brightest outer image, although lens models put the brightness flux ratio at $\sim 0.75$.  In the case of a naked cusp, the two outer images have the same image parity.

\begin{figure}
\includegraphics[width=1.65in]{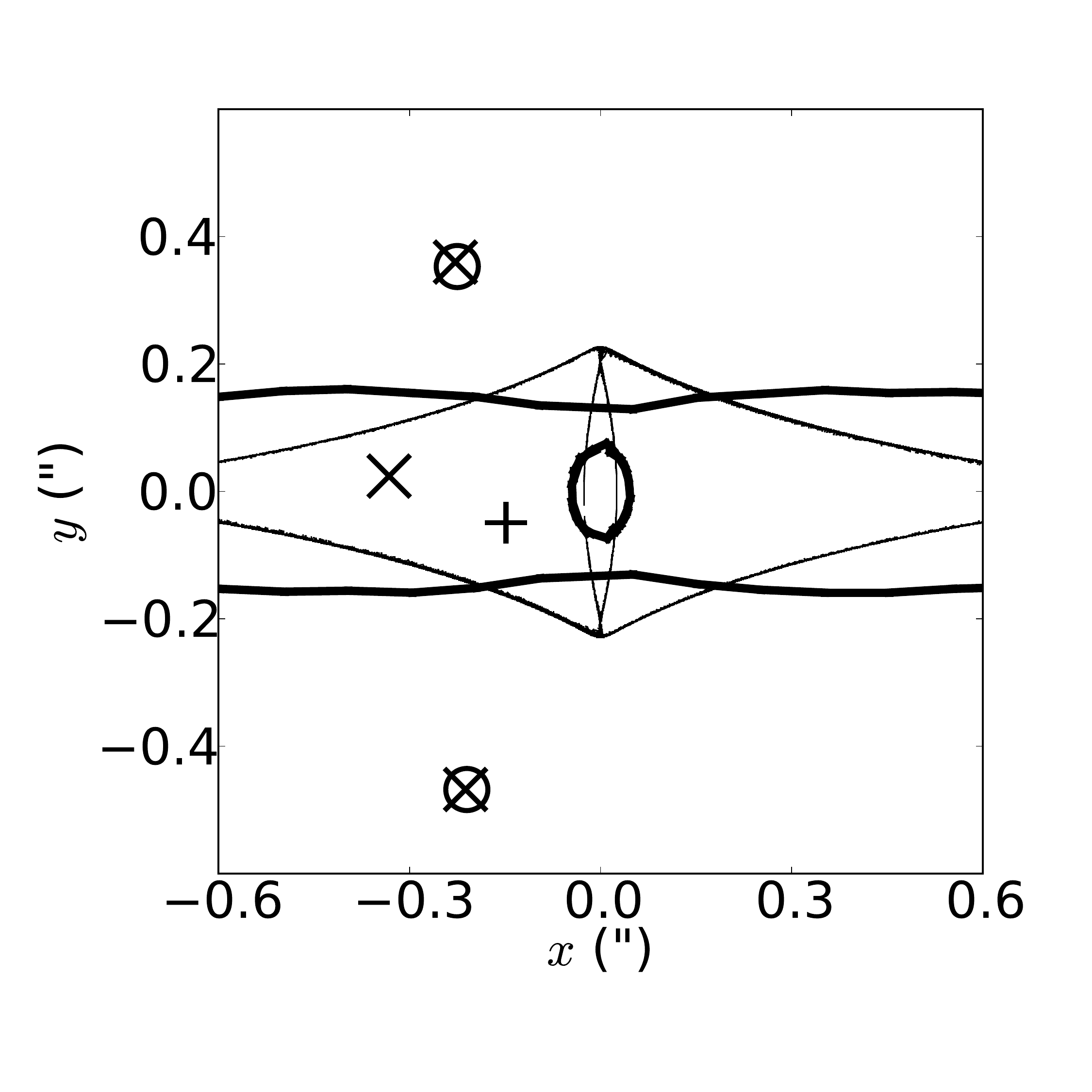} 
\includegraphics[width=1.65in]{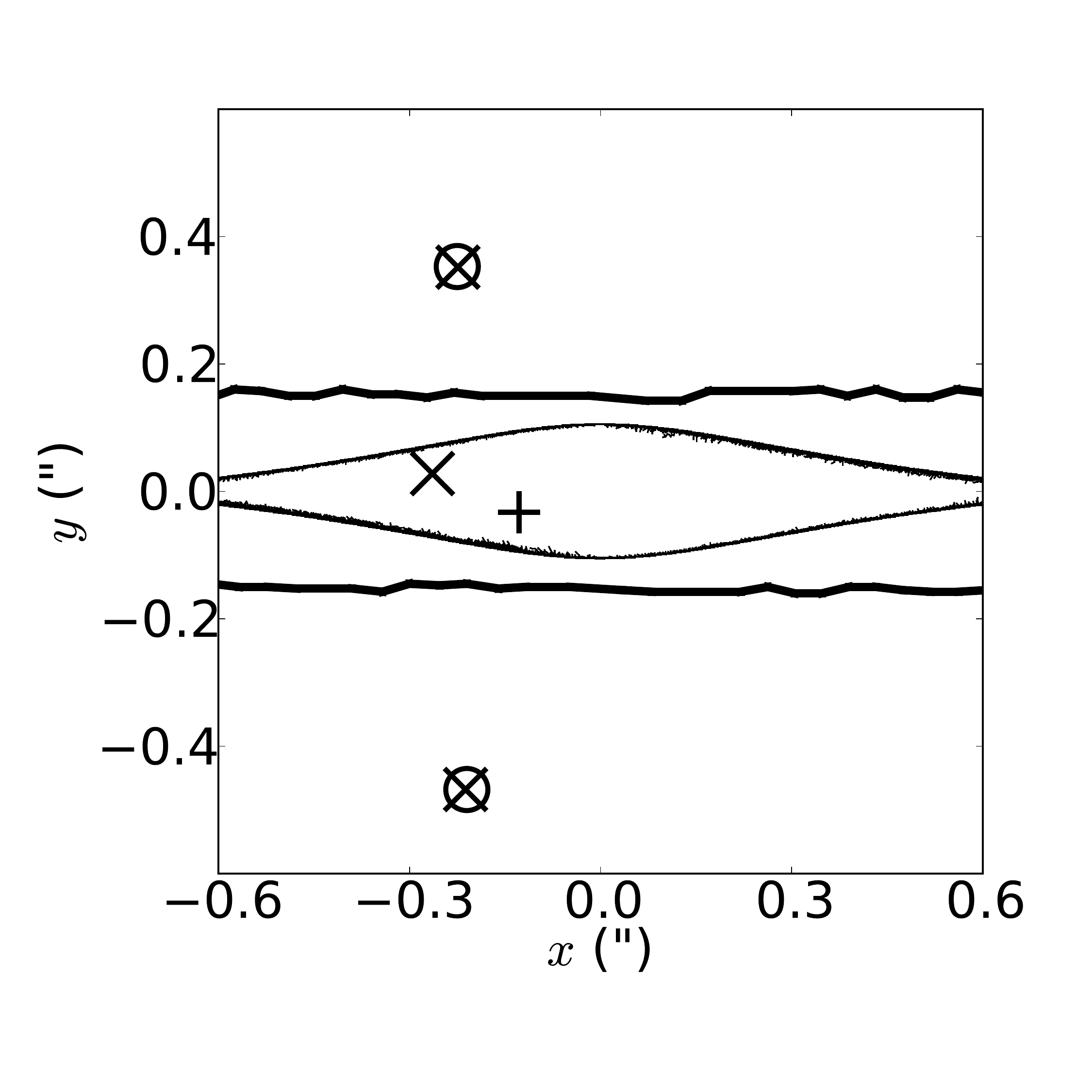} 
\includegraphics[width=1.65in]{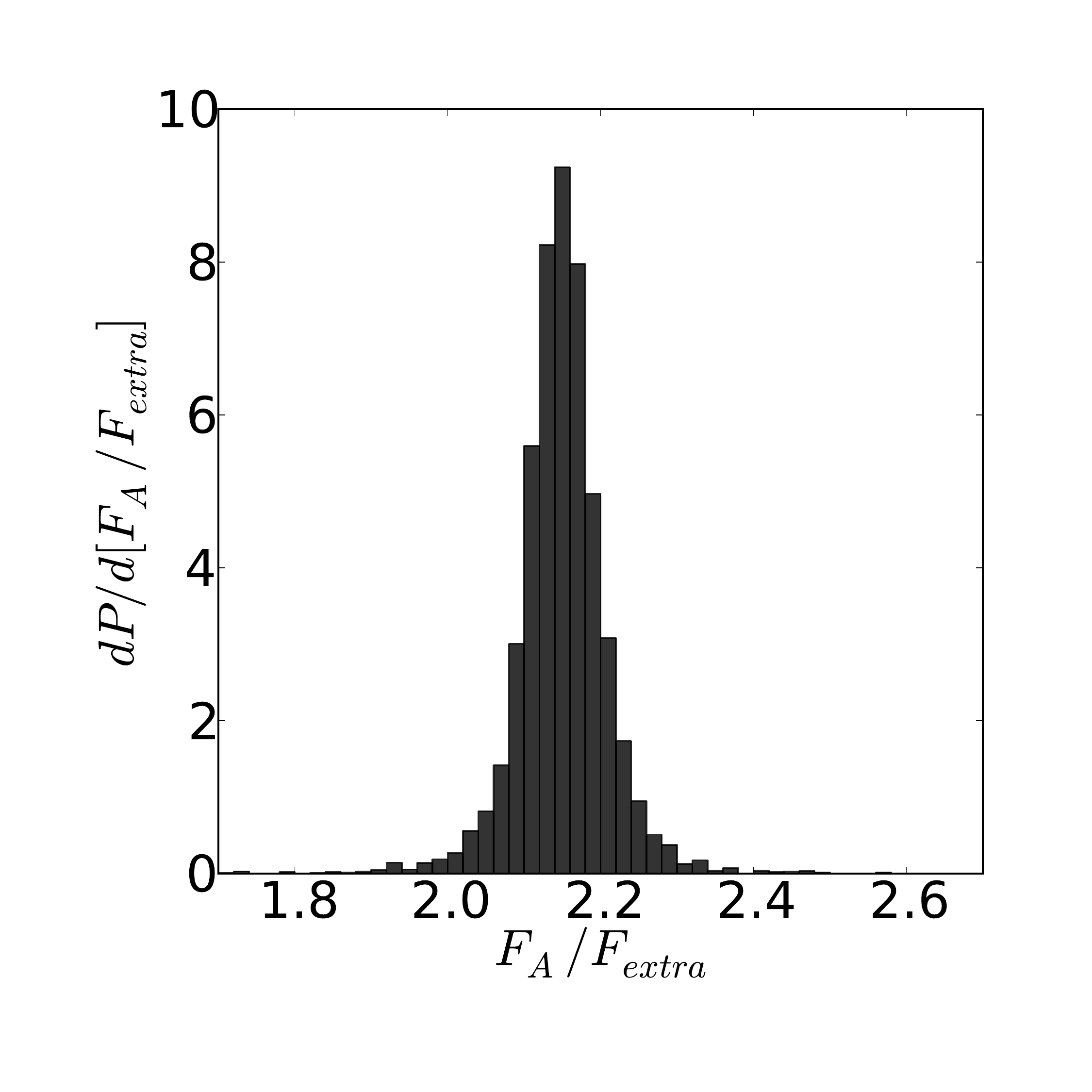} 
\includegraphics[width=1.65in]{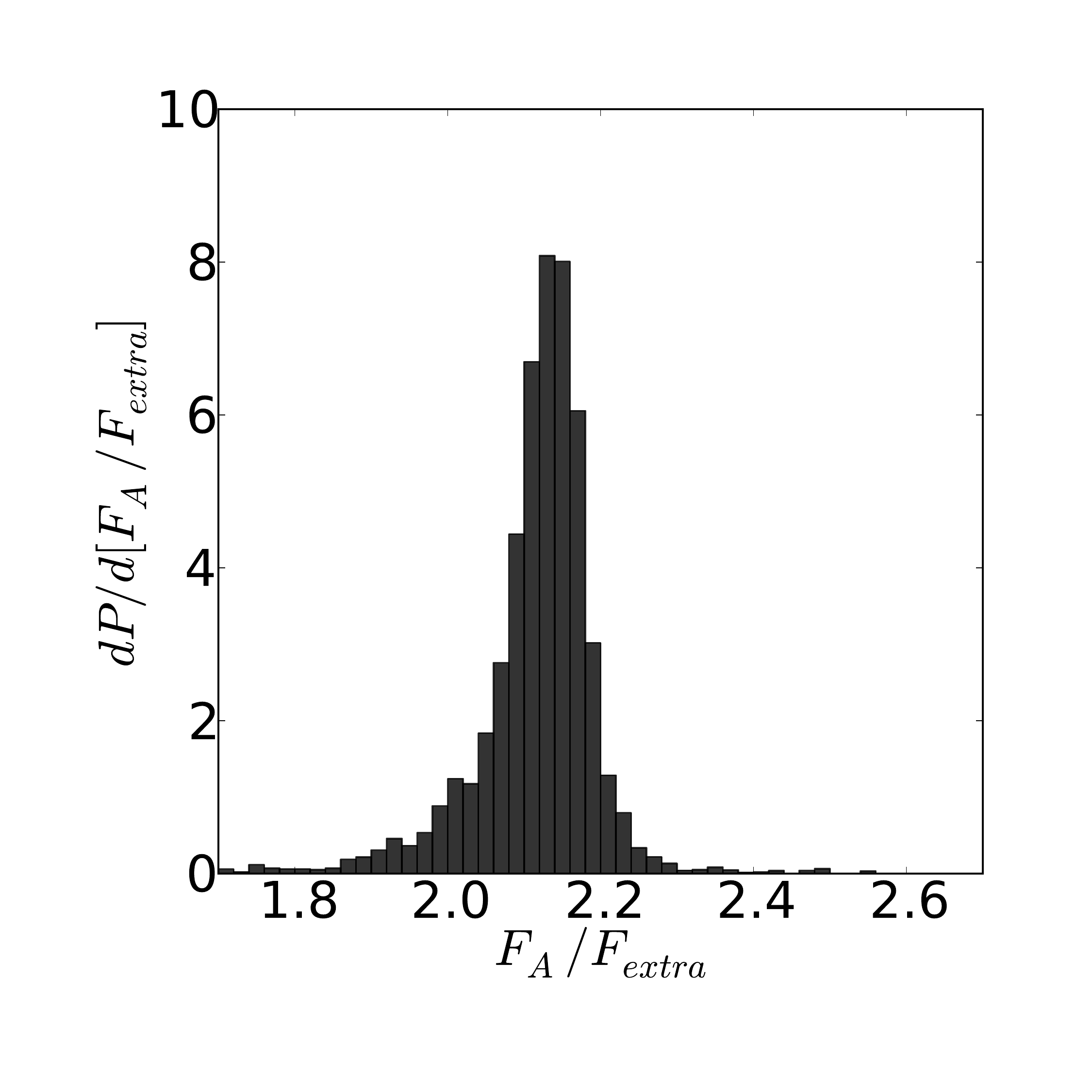} 
\caption{Top:  The critical curve (thick line) and caustic (thin line) of the best-fit model, along with the positions of the source (plus marker) and the images (observed: circles, modeled: crosses).  Bottom:  The marginalized probability for the flux ratio of image A to the extra image.  The left-hand panels show the results using no dust extinction;  the right-hand panels the results using 2 magnitudes of dust extinction. \label{fig:bestfit}}
\end{figure}

\begin{table}
\begin{center}
\caption{Best-Fit Parameter Values for Model with One Extra Image and 2 Magnitudes of Dust Extinction} \label{tab:bestnaked}
\begin{tabular}{lrc}
\hline
\hline
Parameter & Best-fit & 68\% error interval\\ 
& & (unmarginalized)\\
\hline
 $b_h$ ($\arcsec$)&  $0.91$ &$[0,8.7]$ \\
 $q_h$ & $0.55$ & $[0,1]$\\
 $s_h$ ($\arcsec$)&  $1.06$ & $[0,27] $\\
 $b_g$ ($\arcsec$)&  $1.37$ & $[0.71, 2.78]$\\
 $x_s$ ($\arcsec$)&  $-0.13$  &$[-0.17, -0.09]$\\
 $y_s$ ($\arcsec$)& $-0.03$ &$[-0.07, -0.02]$\\
\hline
\end{tabular}
\end{center}
\end{table}

\begin{figure}
\includegraphics[width=3.3in]{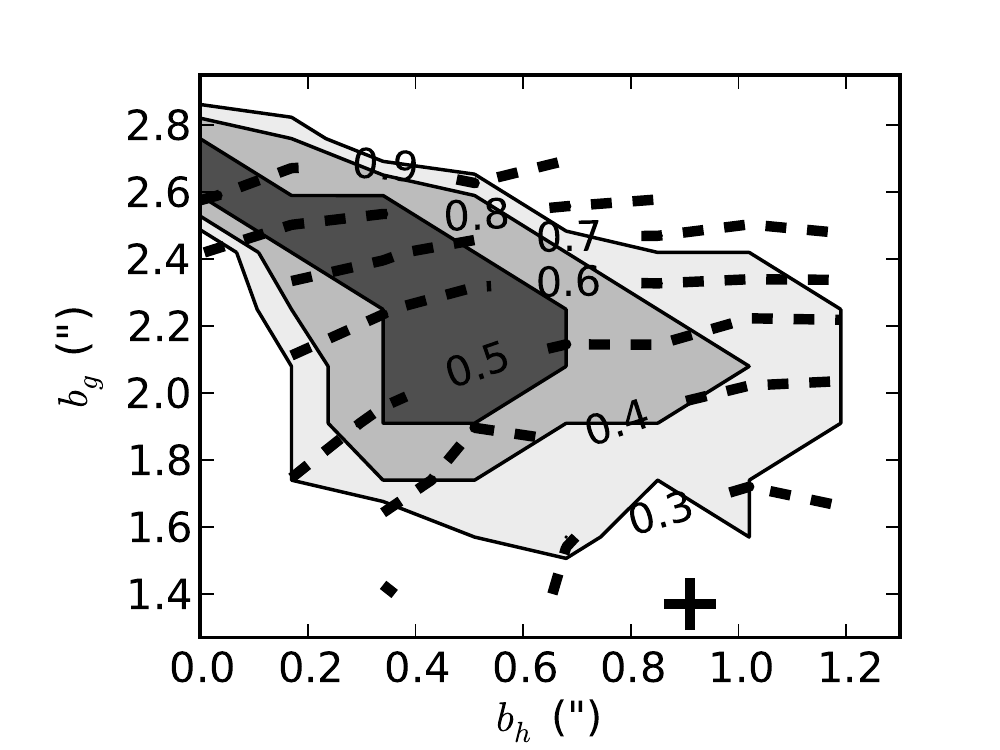} 
\includegraphics[width=3.3in]{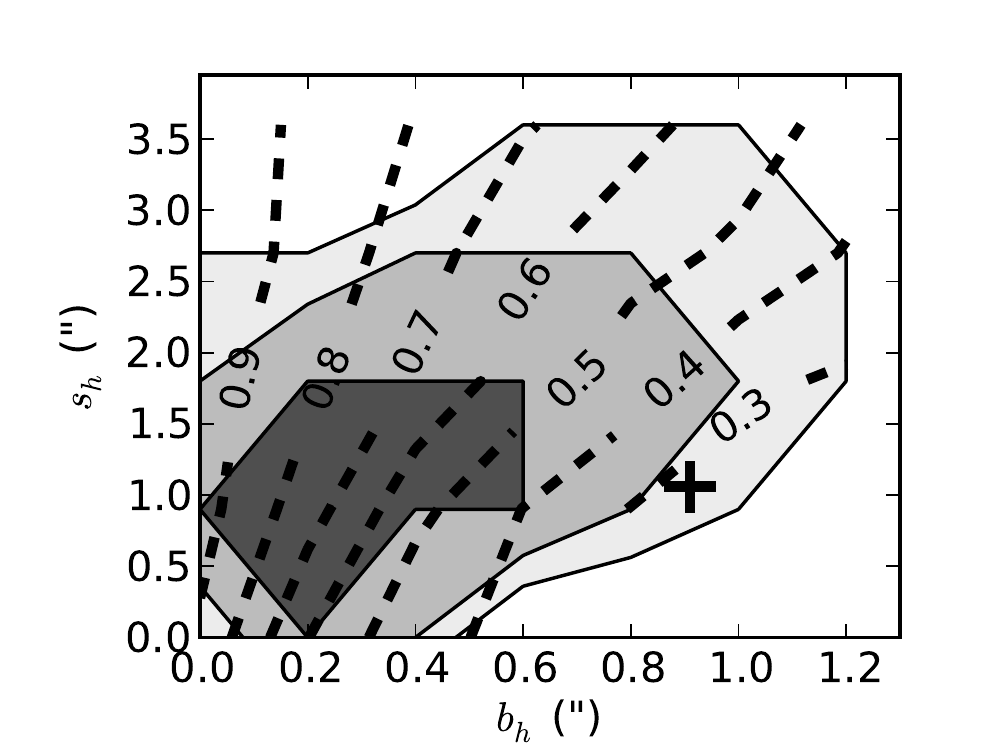} 
\caption{The 80\%, 68\%, and 50\% contours for the the two-dimensional marginalized probability for a model with 2 magnitudes of dust extinction and one extra image.  The top panel shows that the amplitude of the halo and the disk, $b_h$ and $b_g$, are negatively correlated to each other.  The bottom panel shows that $b_h$ is correlated with the size of the core radius in the halo, $s_h$.  The disk mass fractions are shown in labeled dashed contours;  the contours shown are $f_{disk}(2.2R_d) = [0.3, 0.4, 0.5, 0.6, 0.7, 0.8, 0.9]$.  The location of the best-fit parameter value is marked by a plus marker.
\label{fig:degeneracies}}
\end{figure}

A summary of the modeling results using no dust extinction and 1 and 2 magnitudes of dust extinction is found in Table \ref{tab:results}.  Here we see that the no extinction case results in a poorly-fit model and  $\sim 2$ magnitudes of dust extinction is necessary for the modeled image to be faint enough to escape detection.  The best-fit model with one extra image with no dust extinction and the best-fit model with one extra image and 2 magnitudes of dust extinction are described in Figure \ref{fig:bestfit}, which shows the locations of the source, the images, the critical curve (in the image plane), and the caustic (in the source plane).  While the caustics in the two best-fit models are different, both are naked cusp solutions.  In fact, in every case tested, the only acceptable models of the 2 observed images in CX 2201 consist of naked cusp solutions or require a lopsided galaxy.  

We might expect that the model would adapt to the different constraints on the extra image and that the flux of the modeled extra image would be dimmer in the case of no dust extinction than in the case with 2 magnitudes of dust extinction.  This is not the case;  the model does not have the flexibility to make the extra image very dim while also matching the observed images.  Thus, extinction does not play a strong role in the modeling, aside from hiding images that have not been observed.  The probability distribution the flux of the extra image, as shown in Figure \ref{fig:bestfit}, remains the same regardless of the amount of extinction;  the extra image has a flux ratio A/[extra image] $\sim 1.9-2.4$;  and models with more dust extinction have better fits.  We describe the best-fit parameters below using the 2 magnitudes of dust extinction results as our fiducial result.

The best-fit model and 68\% unmarginalized errors for the model with 2 magnitudes of dust extinction are shown in Table \ref{tab:bestnaked}.  
We display unmarginalized errors as they better represent the range of acceptable values when we consider the full set of parameters (and not any subsets of parameters).  The error intervals for the parameters are large and the modeling does not robustly constrain the parameter values.  For example, the entire permitted range of values for $q$, the halo axis ratio, falls within the unmarginalized 68\% error interval, and the probability distribution of $q$ when marginalizing over all the other parameters is flat.  The poor constraints on parameter values are due in part to degeneracies between parameters in the model.  These degeneracies are illustrated in Figure \ref{fig:degeneracies}.  Parameter degeneracies are best represented by a probability distribution for two selected parameters and marginalizing over the remaining parameters.  Marginalized distributions are narrower than the unmarginalized probability distribution cited in Table \ref{tab:bestnaked}.  In addition, given the degeneracies, the complicated parameter space, and the wide unmarginalized errors, the peak of the marginalized distribution will not necessarily match the best-fit value of the parameter.  The two-dimensional marginalized probability show that the amplitude of the halo and the disk, $b_h$ and $b_g$, are anti-correlated with each other.  In addition, $b_h$ is correlated with the size of the core radius in the halo, $s_h$, such that larger $b_h$ is associated with larger $s_h$.

\begin{figure}
\includegraphics[width=3.3in]{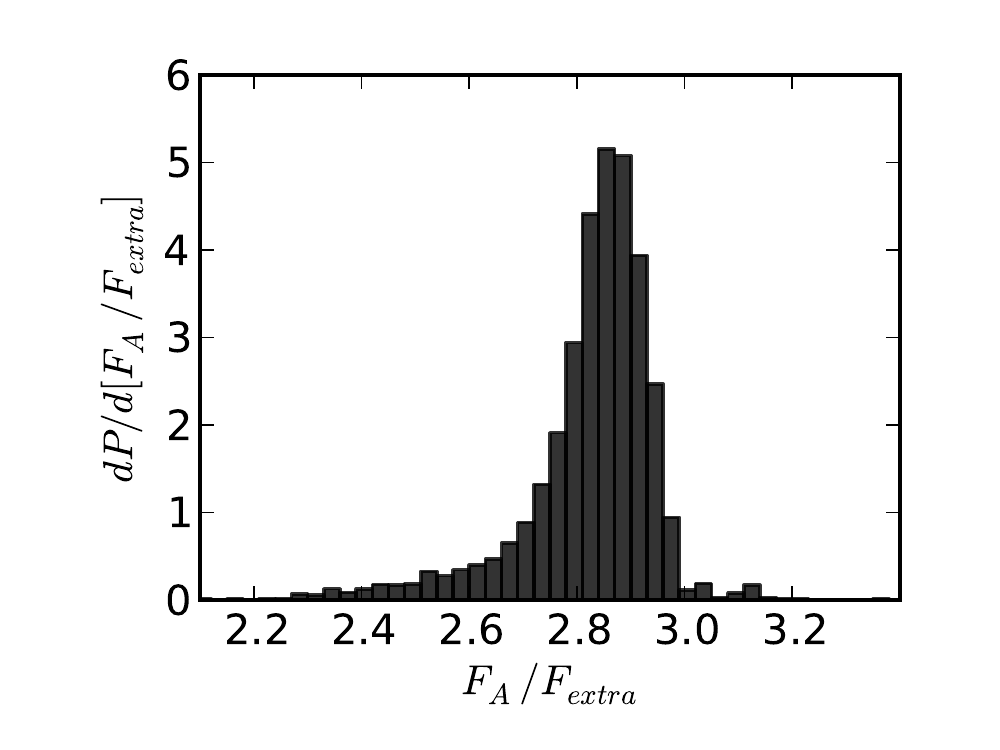} 
\caption{The marginalized probability for the flux ratio A/[extra image] for a model with one extra image, 2 magnitudes of dust extinction, and a smaller disk scale length, $R_d=1\farcs 01$.
\label{fig:rdcomp}}
\end{figure}

In our models we use the mean measured scale length, $R_d = 1\farcs 60$, but the measured values vary with wavelength between $R_d (\arcsec)= [1.01-2.37]$.  We have also tested a smaller value of $R_d=1\farcs 01$.  In general, employing a different $R_d$ doesn't appreciably change the best-fit parameters (not surprising given the size of the unmarginalized errors), but it does have a significant effect on the brightness of the extra image found.  As shown in Figure \ref{fig:rdcomp}, smaller disk scale lengths are associated with dimmer extra images.  

\begin{figure}
\includegraphics[width=3.3in]{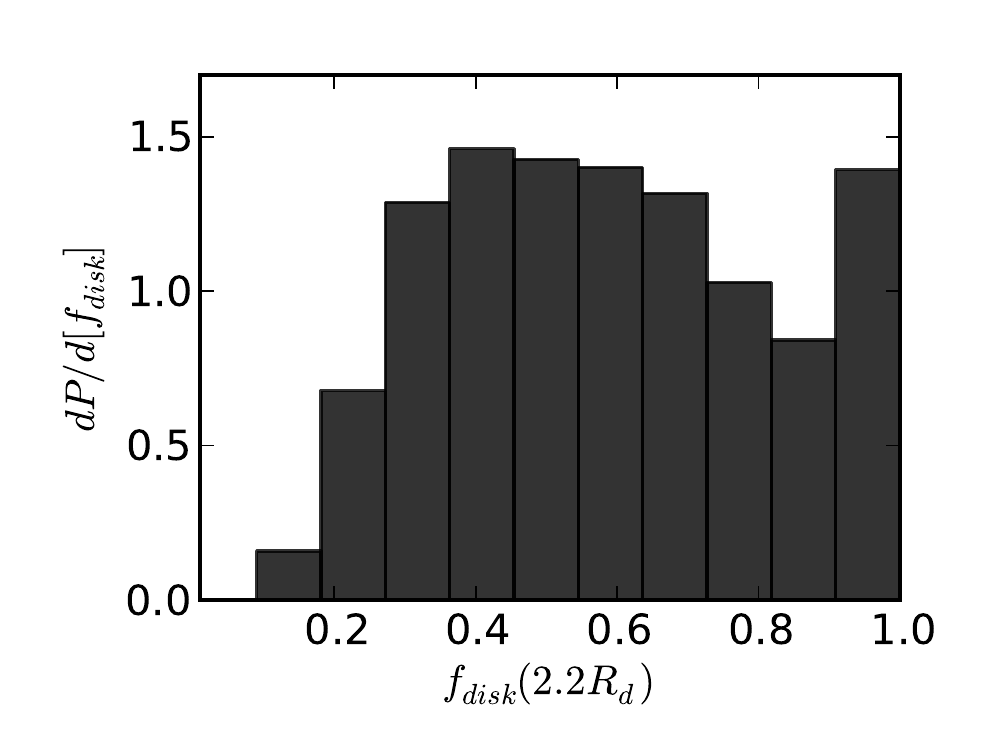} 
\caption{The probability distribution of the fraction of mass in the disk at $2.2R_d$ for the fiducial model with one extra image and 2 magnitudes of dust extinction.  \label{fig:fdm}}
\end{figure}

\begin{figure}
\includegraphics[width=3.3in]{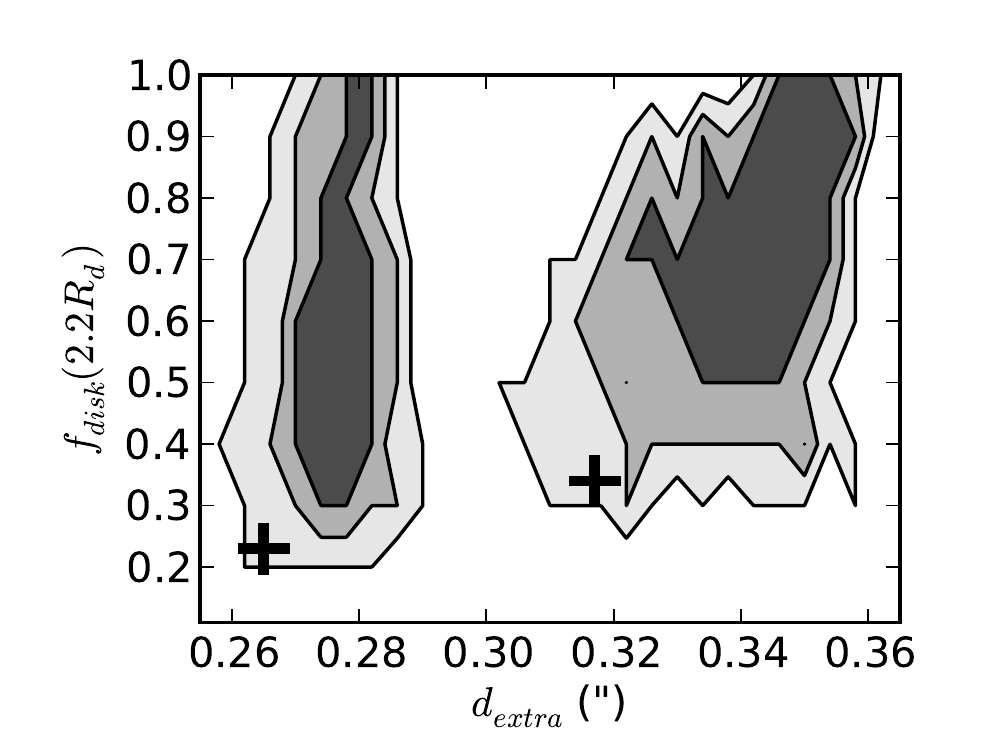} 
\caption{The marginalized probability distribution for $f_{disk} (2.2R_d)$ as a function of the distance of the extra image from the galaxy center.  The contours show where the 85\%, 68\%, and 50\% of the probability lie and the location of the best-fit parameter value is marked by a plus marker.  The set of the contours on the left employ the fiducial disk scale length, $R_d=1\farcs 60$, while the contours on the right use a smaller disk scale length.  $R_d=1\farcs 01$.
\label{fig:dex}}
\end{figure}

We measure the fraction of mass in the galaxy disk  enclosed within a sphere of radius of  2.2 times the disk scale length ($f_{disk}(2.2R_d)$).   The model degeneracies make it difficult to constrain parameters, but the probability distribution of the galaxy disk mass fraction is not flat.  Figure \ref{fig:degeneracies} shows that the disk mass fraction is strongly correlated with the parameter that measures the mass in the galaxy disk, $b_g$, and weakly correlated with the Einstein radius of the halo, $b_h$, because the mass in the dark matter halo within $2.2R_d$ is dependent upon both $b_h$ and the core radius size, $s_h$.  The value of $f_{disk}$ for the best-fit $\chi^2 $ parameter values is $f_{disk} (2.2R_d) = 0.23$,  indicating a sub-maximal disk mass.  As we see from Table \ref{tab:bestnaked}, the errors on the best-fit parameters are large, so the constraints on $f_{disk}$ will be poor.  In Figure \ref{fig:fdm}, we show the probability distribution of $f_{disk}(2.2R_d)$.  The distribution is very wide and only very small fractions ($< 10\%$) are ruled out.  As discussed previously, the best-fit value of $f_{disk}$ is not required to and does not match the peak of the marginalized distribution.  Without additional observational information to break the parameter degeneracies, a definitive statement about the size of the galaxy disk -- whether it is maximal or sub-maximal -- cannot be made.  

Figure \ref{fig:dex} shows the extent to which information about the third image can be used to discriminate between the model fits.  Extra images which are located further from the galaxy center are associated with larger $f_{disk}(2.2R_d)$.  The location of the extra image, however, is also highly dependent upon the disk scale length, such that smaller disk scale lengths are associated with images located further from galaxy center.  Constraining the model with information about the extra image will require breaking the ambiguity surrounding the disk scale length.  Now let us consider the case of a smaller disk scale length, $R_d=1\farcs 01$.  If the extra image is located at $d_{\rm extra} = 0\farcs33$ away from the galaxy center, then the disk mass fraction is $f_{disk} (2.2R_d) = [0.6-0.7]\arcsec$ within 50\% confidence intervals.  If the extra image is located just 20 miliarcseconds (mas) further, $d_{\rm extra} = 0\farcs35$, then 50\% confidence interval expands all the way to $f_{disk} (2.2R_d) = [0.6-1.0]\arcsec$.  So milliarcsecond constraints on the position of the extra image could provide significant constraints on the disk mass fraction.  

The probability distribution of $f_{disk}(2.2R_d)$ for the smaller disk scale length is tilted to larger values than for the fiducial disk scale length.  Fractions smaller than $f_{disk} (2.2R_d) = 20\%$ are ruled out, and, for the best-fit $\chi^2 $ parameter values, $f_{disk} (2.2R_d) = 0.34$.

\subsection{Models with Mulitple Extra Images}

\begin{figure}
\includegraphics[width=1.65in]{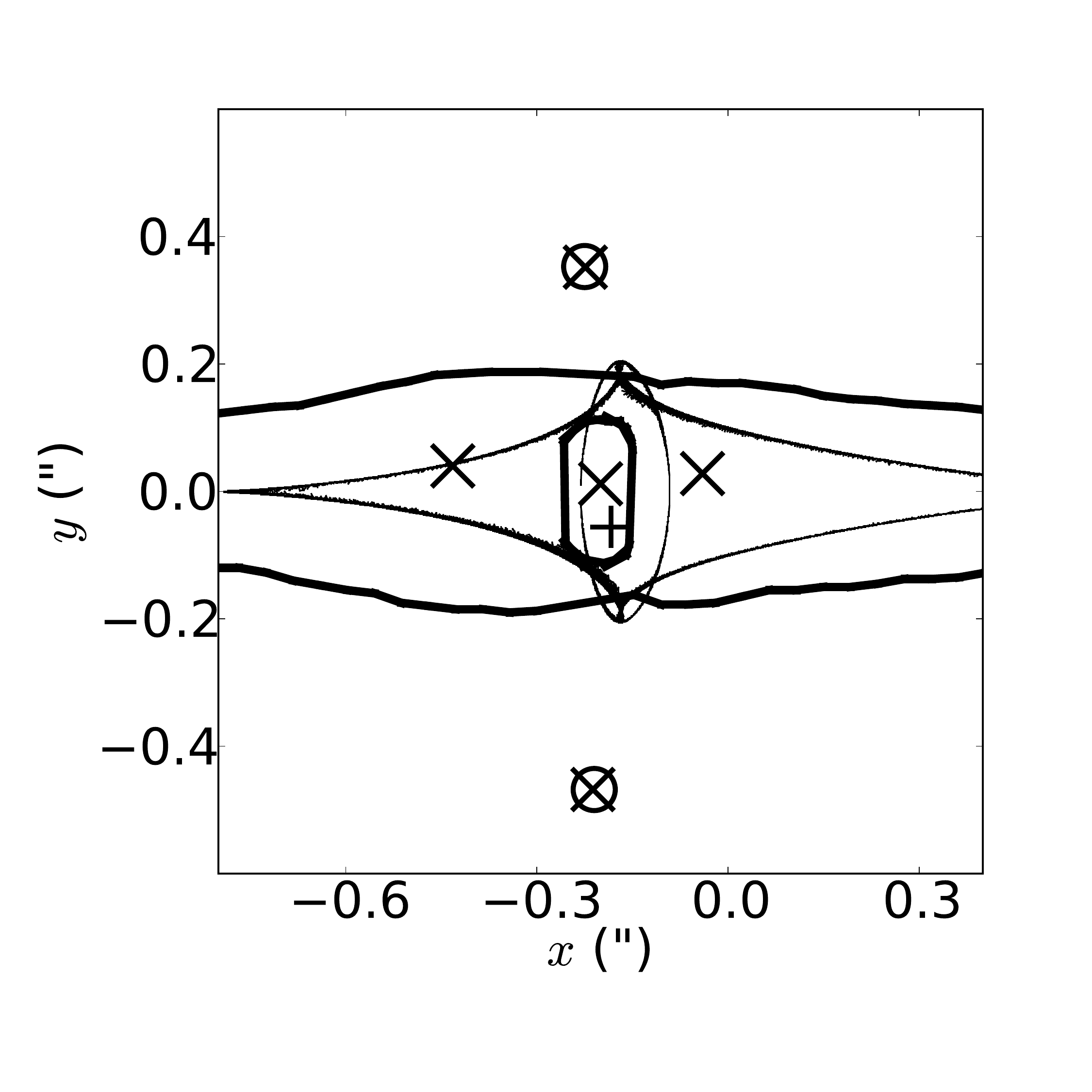} 
\includegraphics[width=1.65in]{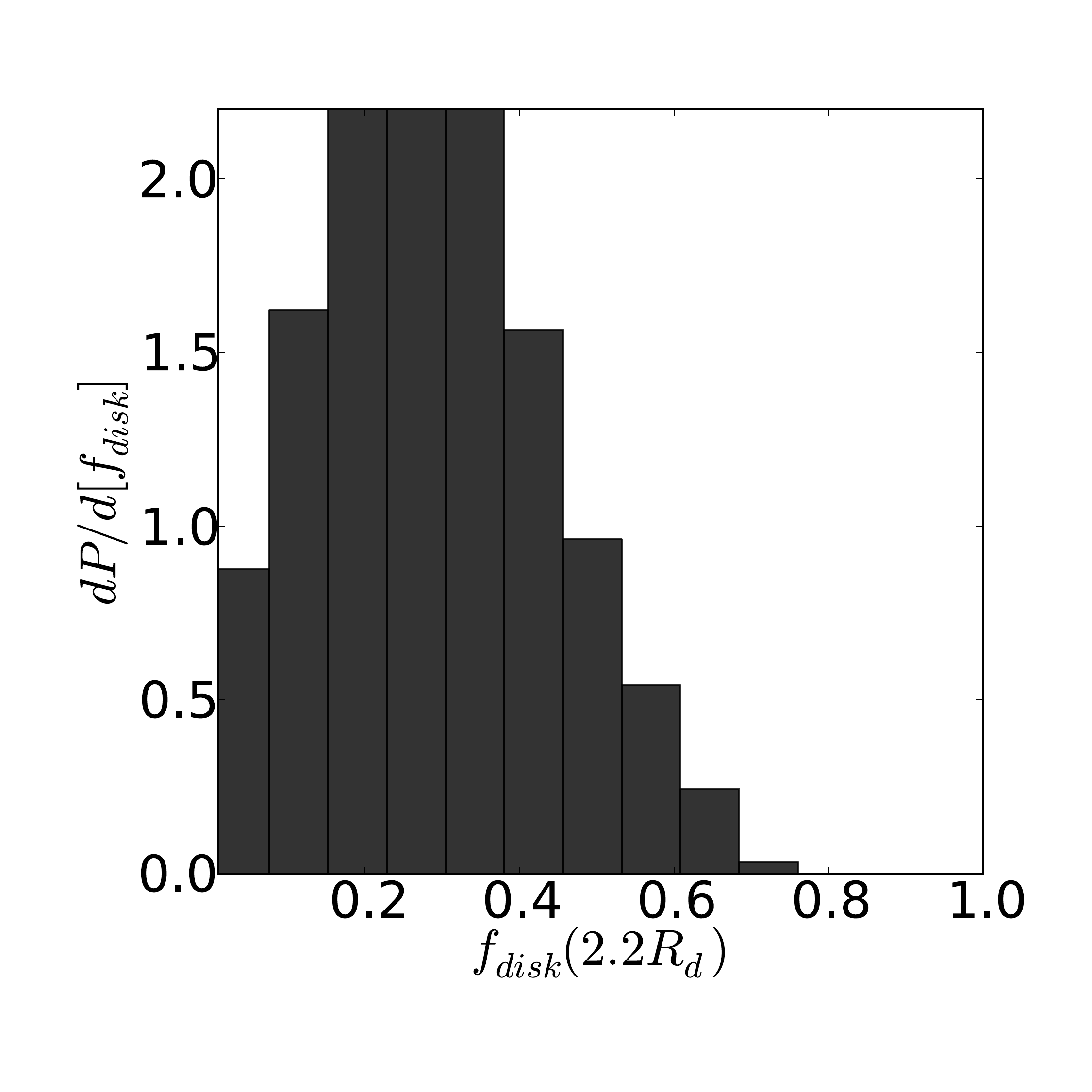} 
\caption{Left:  the critical curve (thick line) and caustic (thin line) of the best-fit model with three extra images, a dark matter halo offset, and 2 magnitudes of dust extinction, along with the positions of the source (plus marker) and the images (observed: circles, modeled: crosses).  Right:  the corresponding probability distribution of the fraction of mass in the dark matter at $2.2R_d$. 
\label{fig:5im}}
\end{figure}

The parameter space for five-image lens systems also consists of naked cusp systems, although the image configuration is no different from 5-image lenses with buried cusps.  In this case, the tangential caustic extends outside of the radial caustic, and the source is located within both caustics.  In general, the five image solution has unobserved extra images which lie along the plane of the disk.  Figure \ref{fig:5im} illustrates this image configuration.  In all cases tested, at least one of the extra images is as bright as the extra image in the 3 image configurations.  Thus, 2 magnitudes of dust extinction are necessary in order to obscure all the extra images.  Regardless the amount of dust extinction the modeled positions of the observed images (not the flux from the three extra images) is the leading factor in the size of the $\chi^2$, and $\chi_{\rm 2im} = 7$, a significantly poorer fit than the best-fit 3-image solution.  

Good 5-image solutions are difficult to produce because no mass configurations exist where two bright images line up perpendicular to the plane of the disk and offset from the galaxy center.  The $x$ position of the observed images are $x = -0.225$ and $x=-0.210$.  The best-fit model positions are $x = -0.232$ and $x=-0.200$;  the model tilts the image positions, bringing one closer to the galaxy center and one further away.  Despite the poorer fit to the model, a five-image solution might be preferable because very flattened dark matter halos are ruled out.  

One option to find a better fit is to, once again, offset the position of dark matter halo.  In our two-component lens model, we can move the center of the dark matter halo while leaving the disk galaxy center at its observed values.  This may be more plausible than shifting both halo and disk to lie collinearly with the lensed images, especially as the size of the offset required is small, $\sim 1$ kpc.  

Offsets between the halo center and the brightest cluster galaxy have been observed in galaxy clusters \citep{allen98,shan_etal10} and in galaxy groups \citep{vandenbosch_etal05b,skibba_etal11}.  In a sample of galaxies from the \ion{H}{1} Nearby Galaxy Survey (THINGS), the offset between galaxy and halo center was less than one radio beam ($\sim10\arcsec$ or 150 $-$ 700 pc) for 13 of 15 galaxies, with the two remaining galaxies exhibiting larger offsets \citep{trachternach_etal08}.  

Evidence for a small offset is also present in our Galaxy.  \citet{su_finkbeiner12} examine gamma-ray emission which may be indicative of dark matter and suggest that there is a $1.5\degr$ (0.2 kpc) offset between the center of the galaxy and the center of the dark matter halo in the Milky Way.  \citet{kuhlen_etal12} use a dark matter plus hydrodynamics simulation of the Milky Way ("Eris") to show that the dark matter halo may be offset from the galaxy center by 300 $-$ 400 pc, and this offset lies preferentially near the plane of the disk.  The predicted offset is smaller than necessary in the CX 2201 system, but it is oriented similarly.  In addition, the \citet{kuhlen_etal12} produced the offset without requiring an external perturber like a passing satellite galaxy or an incomplete accretion event which might be visible in observations of CX 2201.  They suggest that the offset is produced via a density-wave-like excitation by the stellar bar.  

\begin{table}
\begin{center}
\caption{Best-Fit Parameter Values for Model with Three Extra Images, Dark Matter Halo Offset, and 2 Magnitudes of Dust Extinction} \label{tab:best5im}
\begin{tabular}{lrc}
\hline
\hline
Parameter & Best-fit & 68\% error interval\\ 
& & (unmarginalized)\\
\hline
 $b_h$ ($\arcsec$)&  $0.22$ &$[0.08,0.73]$ \\
 $q_h$ & $0.76$ & $[0.17,1.00]$\\
 $s_h$ ($\arcsec$)&  $0.03$ & $[0.00,0.18] $\\
 $b_g$ ($\arcsec$)&  $1.47$ & $[0.00, 2.27]$\\
 $x_s$ ($\arcsec$)&  $-0.18$  &$[-0.22, -0.17]$\\
 $y_s$ ($\arcsec$)& $-0.06$ &$[-0.07, -0.04]$\\
\hline
\end{tabular}
\end{center}
\end{table}

If we offset the center of the dark matter halo to $x_{\rm off}=-0.210$ and $y_{\rm off}= 0.000$ (leaving the galaxy center at the observed location) and with 2 magnitudes of dust extinction, we find a best-fit $\chi^2=0.3$.  The $\chi^2$ from the observed images is as good as that found for the 3-image best-fit solution.  The left-hand panel of Figure \ref{fig:5im} shows the positions of the images, the critical lines and caustics of this best-fit 5-image solution.  We show the best-fit parameter values and the 68\% unmarginalized errors in Table \ref{tab:best5im}.  The probability distribution of $f_{disk} (2.2R_d)$, shown in the right panel of Figure \ref{fig:5im}, rules out very large fractions ($> 80\%$) and disfavors a maximal disk solution.  The best-fit parameter value gives $f_{disk} (2.2R_d) = 0.41$.  While the shapes of the distribution is different than in Figure \ref{fig:fdm}, the constraints on the disk fraction is consistent with the constraints from the 3-image result.

Another option for improving the model fits for 5 image lens systems is to include substructure in the lensing galaxy.   For simplicity, we model the substructure with a pseudo-Jaffe profile \citep[see][]{munoz_etal01}, a profile where the projected surface density is constant inside a core radius, falls as $R^{-1}$ between the core radius and a break radius, and falls as $R^{-3}$ outside the break radius.  If we introduce a substructure clump that is located in the galaxy disk, on the other side of the images from the galaxy center, we can reduce the $\chi^2$ to 0.5.  However, despite several attempts, only massive substructure clumps of $\sim 10^{9} M_{\sun}$ were sufficient to improve the model fit.  In addition, the mass of the substructure must be well concentrated.  We set the break radius to 0.1$\arcsec$ (0.47 kpc) to ensure the mass well contained within the $0.8\arcsec$ separating image A from image B.  A substructure of this mass would be $\sim 10\%$ of the disk galaxy mass.  This size substructure is too large to be plausible if the substructure were to be located near the physical center of the galaxy as it would cause a visible disruption to the galaxy disk.  The substructure could be located far from the galaxy center, as a faint dwarf galaxy hidden behind the disk.  However, the substructure has to be located exactly behind the disk galaxy and further away from galaxy center than images A and B in projection.  Several studies have shown that finding substructure of mass greater than $10^{-5}$ the mass of the lens projected anywhere in the vicinity of the lens center has a probability of only $10^{-3}$ to $10^{-2}$ \citep[e.g.,][]{chen09,xu_etal09,chen_etal11}.  The likelihood of a substructure of $10^9 M_{\sun}$ being serendipitously located near the image positions is small.

\section{Discussion and Conclusions}
\label{sec:conclusions}

Edge-on disk galaxies have the potential to disentangle the relative contributions to the mass from the disk galaxy and the dark matter halo, as the geometry of the luminous disk galaxy differs significantly from the dark matter halo in which the galaxy is embedded.  However, to date, only a small number of edge-on spiral galaxy lenses have been available and have been studied.  

CX 2201 represents a particularly interesting case.  Consisting of a background quasar lensed by an edge-on spiral galaxy into two bright images, the lensed images are of similar brightness and lie to one side of the galaxy center.  Previous efforts to constrain the mass distribution in the galaxy have suggested that at least one additional image must be present \citep{castander_etal06}.  These extra images may be hidden behind the disk which features a prominent dust lane.  The image configuration is also suggestive of a naked cusp configuration, of which only one galaxy-scale lens system has been observed to date.  

\subsection{Summary of Results}

We present Hubble Space Telescope (HST) observations of the system, better constraining the observable parameters of the lens system but without detecting any extra images.  We have explored a range of models to describe the mass distribution in the system, discovering that a variety of acceptable model fits exist.  Our conclusions are summarized as follows:

\begin{enumerate}

\item 2-image lensing solutions require that the galaxy center lie between the two observed images -- a constraint not supported by current observational evidence which puts the galaxy center $0.2\arcsec$ (1 kpc) away.

\item Models which create extra images require at least one image have a flux ratio (image A to extra image) of $\sim 2-3$.  This is sufficiently bright to be detectable, and 2 magnitudes of dust extinction is necessary in order to obscure extra images.

\item  All likely solutions with 3 images require a naked cusp.  However, the parameter space is not well constrained.  Given the degeneracies in the model parameters, it is difficult to put bounds on the fraction of mass in the disk within 2.2 times the disk scale length.  Very small fractions ($< 10\%$) are ruled out, and the best-fit model finds $f_{disk} (2.2R_d) = 0.23$.  An observational constraint on the unobserved extra image would break many of the parameter degeneracies and put strict limits on the dark matter halo fraction.  

\item The observed images are poorly fit by a 5-image solution, unless the dark matter halo center is moved to lie between the two images.  Such an offset might be well-motivated by simulations of the Milky Way which show a similar offset.  Large disk mass fractions ($> 80\%$) are ruled out and the best-fit model finds $f_{disk} (2.2R_d) = 0.41$.  Given the poor constraints, the disk mass fractions are consistent with the 3-image result. 

\end{enumerate}

\subsection{Future Observations with JWST}

In order to put robust constraints on the mass distribution of the lensing galaxy, better constraints on the disk scale length and imaging of the extra image or images in CX 2201 are needed.  In both cases, observations at longer wavelengths are the key to reducing the impact of dust extinction.  The James Webb Space Telescope (JWST) is expected to observe the near and mid-infrared ($0.6 - 30\mu$m) with unprecedented resolution and sensitivity.  In addition to longer wavelengths, JWST observations will provide further improvements as it will feature a smaller PSF than current HST observations.

We estimate the amount of dust extinction that may be observed in CX 2201 at JWST wavelengths using the fitting formula given by  \citet{calzetti97} and \citet{calzetti_etal00}.  The longest HST wavelength observation we use is at $1.6\mu$m $H$ band ($1.2\mu$m in the  rest frame).  At this wavelength our extra images fall below the threshold of detection and at least 2 magnitudes of dust extinction are required.  Let us consider the case where there are exactly 2 magnitudes of dust extinction in the $H$-band.  If we observe CX 2201 in $K$-band ($2.2\mu$m), there are $1.1$ magnitudes of extinction and extra images with flux ratios (A/[extra image]) smaller than 3.6 are detectable.  If there are 3 magnitudes of dust extinction in the $H$-band, then at $K$-band there are $1.6$ magnitudes of extinction and extra images with flux ratios smaller than 2.3 are detectable.  In both these cases, the extra images in our fiducial models would be observable (c.f., Figure \ref{fig:bestfit}).  The Calzetti relations for dust extinction encompass the far-UV to near-IR spectral range and the longest wavelength probed is $2.2\mu$m.  If we extend the relation to observations in the $L$ band ($3.5\mu$m observed, $2.6\mu$m rest frame), just outside of the Calzetti spectral range, we find that there are $0.25$ and $0.37$ magnitudes of dust extinction for 2 and 3 magnitudes in the $H$ band, respectively, and flux ratios (A/[extra image]) smaller than $\sim 7.5$ are detectable.  In fact, if there were 12 magnitudes of extinction in our $H$ band observations (more than the 8 magnitudes estimated for the Milky Way), we could still detect the extra images predicted by our models in the $L$-band.  

At longer wavelengths, we would expect the amount of dust extinction to be even lower than these limits.  However, at those wavelengths attempts to observe extra images will be hampered by dust emission, and observations of CX 2201 between near-infrared and mid-infrared wavelengths may be where both effects are minimized \citep[see, for example,][]{dacunha_etal08}.  Thus, future observations with the James Webb Space Telescope (JWST) represent the most promising avenue to observing central images and measuring the mass of the disk in CX 2201.

\section*{Acknowledgements}
The authors would like to thank Jeffrey Blackburne for his invaluable assistance with the data reduction.  JC acknowledges support through grant HST: GO:10518.  JM gratefully acknowledges support from CONICYT project BASAL PFB-06.

\bibliography{manuscript}

\end{document}